\documentclass[a4paper]{jpconf}
\usepackage[utf8]{inputenc}
\usepackage{amsfonts}
\usepackage{graphicx}

\def\pbar*{\ensuremath{\bar{\textnormal{p}}}}
\def\Hbar*{\ensuremath{\bar{\textnormal{H}}}}
\def\Hbarplus*{\ensuremath{\bar{\textnormal{H}}^+}}
\def\nps*{\ensuremath{n_{\small\textnormal{Ps}}}}
\def\sps*{\ensuremath{\sigma_{\small\textnormal{Ps}}}}
\def\npb*{\ensuremath{N_{\small\textnormal{\pbar*}}}}
\def\nhb*{\ensuremath{N_{\small\textnormal{\Hbar*}}}}
\def\shb*{\ensuremath{\sigma_{\small\textnormal{\Hbar*}}}}
\def\nhbp*{\ensuremath{N_{\small\textnormal{\Hbarplus*}}}}
\def\shbp*{\ensuremath{\sigma_{\small\textnormal{\Hbarplus*}}}}

\begin{document}

\title{Enhanced anti-hydrogen ion production}
\author{D. A. Cooke$^1$ and A. Husson$^2$ and D. Lunney$^2$ and P. Crivelli$^1$}
\address{$^1$ ETH Zurich, Institute for Particle Physics, 8093 Zurich, Switzerland}
\address{$^2$ CSNSM-IN2P3/CNRS, Universit\'e de Paris Sud, 91405 Orsay,  France}
\ead{crivelli@phys.ethz.ch}

\begin{abstract}
The production of anti-hydrogen ions (\Hbarplus*) in the GBAR experiment will occur via a two step charge exchange process. In a first reaction, the anti-protons from the ELENA ring will capture a positron from a positronium target producing anti-hydrogen (\Hbar*) atoms. Those interacting in the same positronium target will produce in a second step \Hbarplus*. This results in a dependence for the \Hbarplus* production rate which is roughly proportional to the positronium density squared. We present a scheme to increase the anti-ions production rate in the GBAR experiment by tailoring the anti-proton to the positron pulse in order  to maximise the temporal overlap of Ps and \pbar*. Detailed simulations show that an order of magnitude could be gained by bunching the anti-protons from ELENA. In order to avoid losses in their capture in the Paul trap due to the energy spread introduced by the bunching, debunching with a symmetrical inverted pulse can be applied to the \Hbarplus* ions. 
\end{abstract}

\section{Introduction}
Different theoretical explanations have been put forward and experimental efforts are ongoing to try to address the origin of the Baryon Asymmetry observed in our Universe. Those include activities focusing on the gravitational behaviour of antimatter. No compelling theoretical argument seems to support that a difference between the gravitational behaviour of matter and antimatter should be expected \cite{Nieto1991}, although some attempts have been made to show the contrary \cite{Scherk1979, Chardin1992, Chardin1993, Chardin1997} and it would be allowed in the context of the Standard Model Extension (SME) \cite{Kostelecky2011, Kostelecky2016}. Moreover, observations and experiments have been interpreted as evidence against the existence of antigravity type forces \cite{Good1961,Pakvasa1989,Apostolakis1999,Gabrielse1999}. However, those could be argued to be model dependent and, therefore, a simple free fall measurement with a straightforward interpretation is preferable. A first attempt in this direction was made recently by the ALPHA collaboration \cite{ALPHA2013} that bound the ratio of gravitational mass to inertial mass of antihydrogen between $-65$ and 110.\par

The first proposal to measure directly the gravitational force acting on antiparticles using positrons was done 50 years ago by Witteborn and Fairbank \cite{Witteborn1967,Witteborn1968}. Such a measurement was later proposed at CERN with antiprotons \cite{Holzscheiter2014}. However, those experiments have not yet been realised since the use of charged particles to measure the effect of gravity requires the elimination of stray electromagnetic fields to a level which is extremely challenging. The solution to this problem is to use neutral systems composed entirely of antimatter such as antihydrogen \cite{AEGIS2012,GBAR2015,Hamliton2016} or partly such as muonium and positronium \cite{Kirch2014,Cassidy2014,Crivelli2015}.

\section{GBAR experiment}
The Gravitational Behaviour of Antimatter at Rest (GBAR) is an experiment in preparation at CERN \cite{GBAR2011}. Its goal is to measure the gravitational acceleration $\bar{g}$ imparted to freely falling anti-hydrogen (\Hbar*) atoms, in order to perform a direct experimental test of the Weak Equivalence Principle with antimatter. The objective is to reach a relative precision on $\bar{g}$ of 1\% in a first stage, with the perspective to reach a much higher precision ($10^{-4}$--$10^{-5}$) using quantum gravitational states \cite{Dufour2014} in a second stage.\par

The principle of the experiment is described in detail in \cite{GBAR2015, GBAR2011}. Here we review briefly the proposed experimental technique based on the original idea of T. H\"ansch and J. Walz \cite{Walz2004} to first produce \Hbarplus* ions from interactions of anti-protons (\pbar*) with ortho-positronium (Ps) as:
\begin{eqnarray}
& \textnormal{Ps} + \pbar* \to \Hbar* + \textnormal{e}^- \label{eqn:hb}\\
& \textnormal{Ps} + \Hbar* \to \Hbarplus* + \textnormal{e}^- \label{eqn:hbp}
\end{eqnarray}
Reaction \ref{eqn:hb} has been demonstrated for normal hydrogen by Merrison et al. \cite{Merrison1997}. The production of \Hbar* in Rydberg states has already been proven by the ATRAP collaboration via a two step charge exchange, i.e. formation of Rydberg positronium with positrons on Cs atoms and subsequent formation of $\Hbar*^*$ \cite{Speck2004,McConnell2016}. The cross sections were calculated by different authors \cite{Rawlins2016} and those are in good agreement with the available experimental data. The cross sections of reaction \ref{eqn:hbp} were computed \cite{Walters2007,Comini2013} but have not been measured yet.\par

The \Hbarplus* ions produced in GBAR will then be cooled down in two steps: first in a Paul trap and finally in a linear trap to a few tens of $\mu$K through sympathetic cooling with cold beryllium ions \cite{Hilico2014}. A laser pulse will be applied to detach the excess positron so that the free-fall measurement time for the neutral atom can start. The subsequent \Hbar* annihilation on a plate placed at about 10 cm from the trap is detected and provides the stop signal used to measure the free fall time and thereby extract $\bar{g}$. The choice of producing \Hbarplus* ions to get ultracold antihydrogen atom is the specificity of the GBAR experiment.\par

The \Hbarplus* production yield can be estimated with:
\begin{equation}
\nhbp* \approx \npb* \left(\shb*\nps*L\right)\left(\shbp*\nps*L\right) \label{eqn:yield}
\end{equation}
where $\shb*$, $\shbp*$ are the cross sections of reactions 1 and 2, $\nps*$ is a characteristic Ps density, and $L$ a characteristic length of the interaction volume. Ps will be formed by dumping a bunch of few $10^9$ positrons (with $\sigma \approx 10$ ns) onto a porous silica target. Located in a $1\times1\times20$ mm tube, Ps emitted back into vacuum is confined by reflection from the walls to a small volume which the \pbar* beam pass through. Hence, $\nps*$ is a time- and space-varying quantity, so that even in the limit of low interaction probability, Eq. \ref{eqn:yield} is approximate. The evolution of the Ps density inside the tube was simulated using GEANT4 \cite{Crivelli2016}, and the resultant formation probability of \Hbar* by passage of an \pbar* bunch through the evolving cloud calculated, as a function of offset time between Ps and \pbar* pulses. The antiprotons from ELENA \cite{ELENA2014} are coming in bunches of $\sigma = 75$ ns and therefore the temporal overlap is clearly not maximised. As the number of Ps atoms available decays exponentially, compression of the \pbar* bunch to the shortest possible width in order to overlap with the maximum Ps density has a dramatic effect on \Hbar* production.\par 

Here we propose to first bunch the antiprotons in order to increase the \Hbarplus* formation probability, followed by time debunching (bunching in energy) the \Hbarplus* to keep the energy spread below 100 eV in order not to affect the capturing efficiency in the Paul trap. 


\section{Simulation}
\begin{figure}[h!]
\centering
\includegraphics[width=0.8\textwidth]{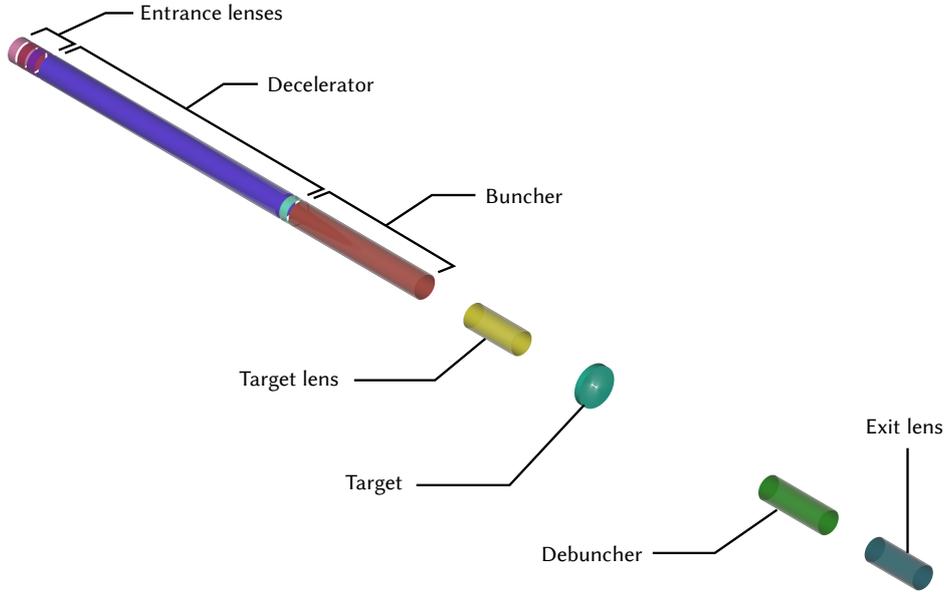}
\caption{Simulation geometry. The complete length of the simulation is 5 m.}
\label{fig:geom}
\end{figure}

The simulation, performed using SIMION 8.0, comprises a five metre beam pipe with decelerator, buncher, and debuncher stages, with additional ion focussing elements. A pulse of antiprotons was created with a Gaussian distribution of kinetic energy with mean 100 keV and $\sigma_{KE} = 0.07$ keV, beamspot size of $\sigma_r = 0.5$ mm and opening angle of 1.5 mrad, and time distribution of $\sigma_{t_0} = 75$ ns, corresponding to typical expected bunch parameters from the ELENA ring. The simulation geometry is depicted in Figure \ref{fig:geom}.

As proposed by D. Lunney for GBAR, deceleration from 100 keV to $\sim 6$ keV is achieved using a voltage step from -94 kV to 0 V applied to a 1 m long cylindrical electrode (diameter 80 mm) while the antiproton buncher is entirely inside it. This is modelled as an exponential decay of the voltage on the tube, with a time constant of 100 ns---allowing an antiproton drift time of approximately 10 times this time constant. Entrance lenses are required to focus the bunch through the decelerator in order to avoid significant beam divergence. Following the decelerator there is a time bunching electrode where a linear ramp potential is applied as the antiproton bunch enters, resulting in larger deceleration for early arriving particle in comparison to late arriving ones. Although an ideal buncher for monoenergetic particles with only a single electrode uses a potential varying as $t^{-2}$ \cite{Mills1980}, a linear ramp is employed here as the energy spread on exit from the decelerator is large. As a result, no significant improvement in bunching performance can be achieved using a $t^{-2}$ ramp, and a high-voltage linear ramp is technically easier to achieve. A single cylindrical lens is used to focus the bunch through the target, with the target itself represented by a 2 mm diameter, 20 mm long hole.\footnote{In order to reduce the memory requirements of SIMION, axial symmetry was preserved throughout.}\par

After the target, the simulated particle type is transformed into \Hbarplus* with the momentum directed in the same way as the incident \pbar* beam, and the bunch passes through an energy bunching (or time debunching) stage. This has an identical voltage ramp to the bunching stage, offset to later times (and with the opposite sign), and works on a similar principle. After time bunching, the \pbar* pulse has a velocity distribution correlated with the time distribution, so that after the time focus, earlier arriving particles have higher velocities than later arriving ones. This means that another time varying potential can undo the time bunching effect, restoring the original energy distribution (in an ideal case). Following the debuncher, a final exit lens is used to focus the beam at the end distance.\par

\begin{figure}[h!]
\centering
\includegraphics[width=0.8\textwidth]{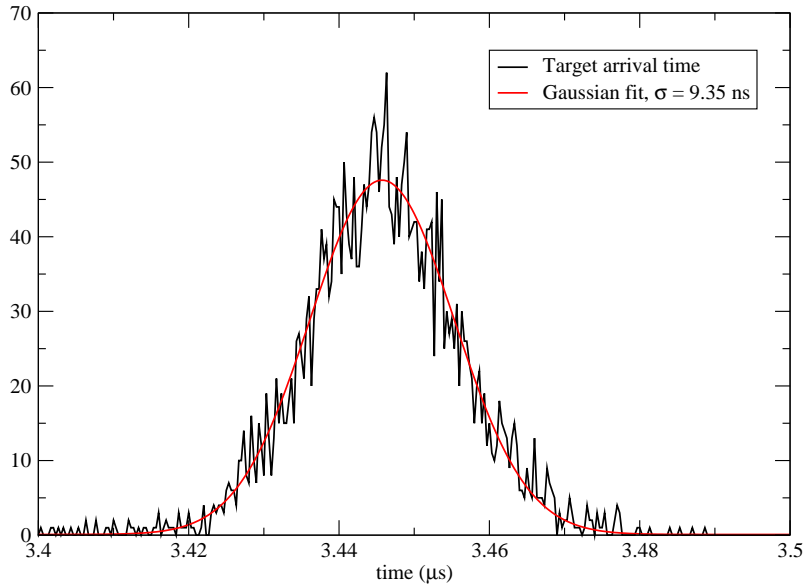}
\caption{Time distribution of antiproton bunch passing through the target aperture.}
\label{fig:time}
\end{figure}

The time distribution at the target is shown in Figure \ref{fig:time}, together with a Gaussian fit to the data, revealing compression in time by a factor of approximately 7.5. Improvement beyond this level is rendered difficult owing to the energy spread of the initial beam. The voltage ramping requirements are a rate of approximately 10 kV $\mu$s$^{-1}$ over a voltage range of nearly 4 kV, which can be achieved using high voltage solid state switches. Focussing at the target allows approximately 35\% of the bunch to pass through a $1\times1$ mm hole in the centre of this (corresponding to a more realistic target geometry).\par

\begin{figure}[h!]
\centering
\includegraphics[width=0.8\textwidth]{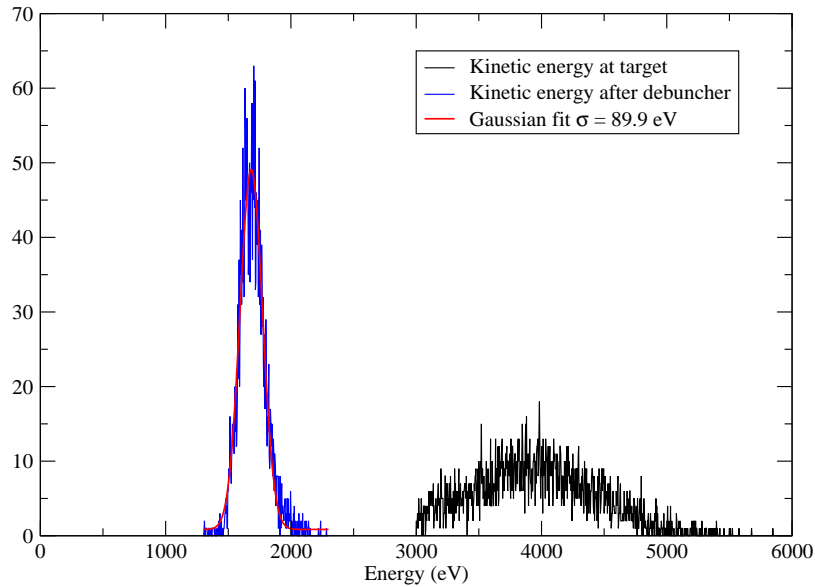}
\caption{Energy distribution of \pbar* bunch at target and \Hbarplus* at simulation end point (equivalent to entrance of quadrupole trap).}
\label{fig:energy}
\end{figure}

\begin{figure}
\centering
\includegraphics[width=0.8\textwidth]{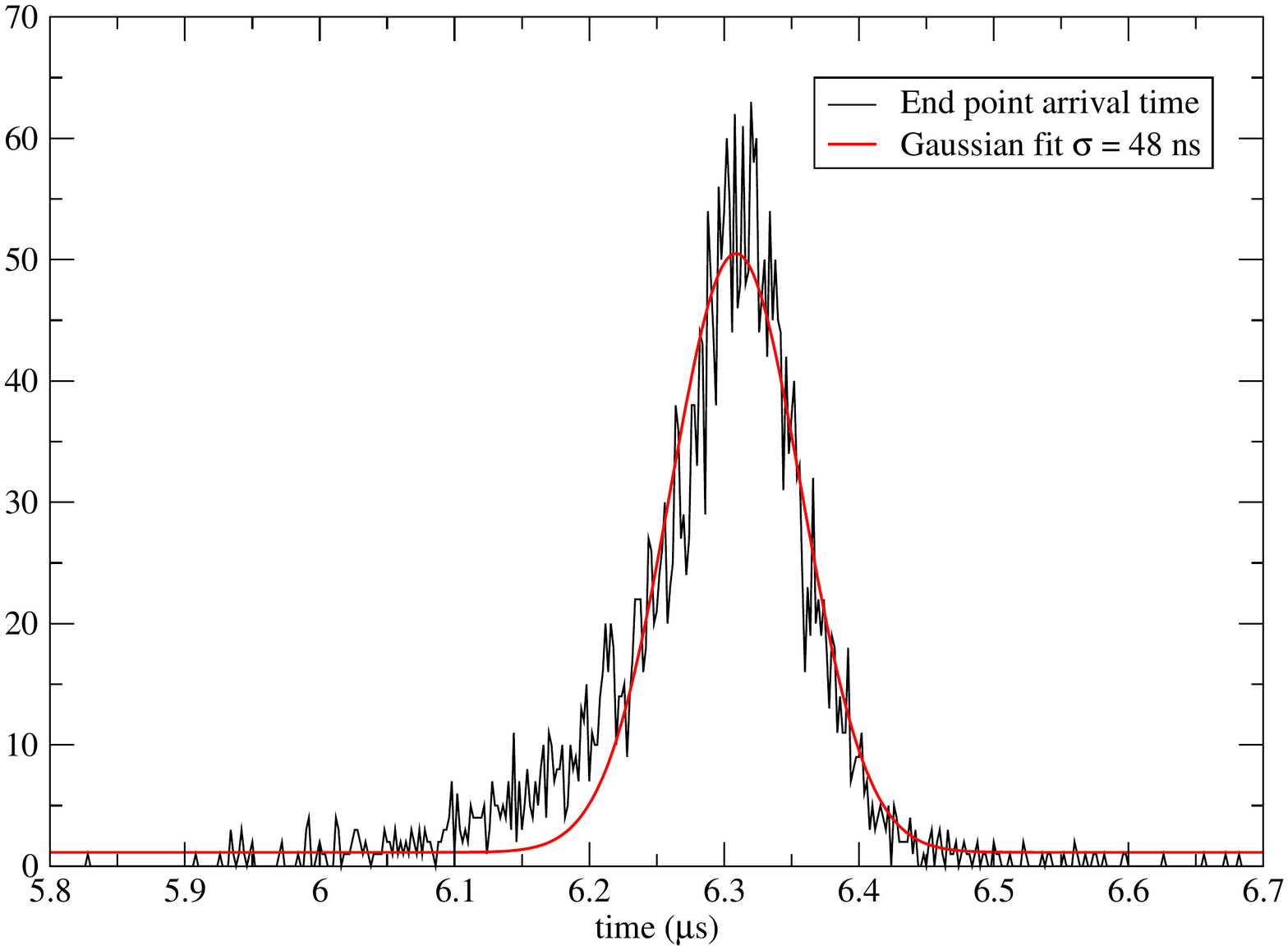}
\caption{Time distribution of antiproton bunch at simulation end point.}
\label{fig:endtime}
\end{figure}

Figure \ref{fig:energy} shows the kinetic energy distribution of the \pbar* bunch at the target position. This is naturally broadened by the time bunching process, which also introduces a shift to lower energies. In principle, the decelerator and buncher parameters can be optimized for arbitrary target energies, so a typical mean value in the range 0--10 keV is presented here. Also shown on Figure \ref{fig:energy} is the kinetic energy distribution at the simulation end point, after the debunching stage. This is almost completely restored to the original energy spread at the input to the decelerator stage, which is the theoretical limit for this process (assuming no particle loss, and that the original energy distribution was uncorrelated with the original timing distribution). Figure \ref{fig:endtime} shows the time distribution for particles reaching the end of the simulation (having passed through the target aperture. The width of $\sigma = 48$ ns is consistent with the requirements for high efficiency capture in a subsequent Paul trap.\par

\begin{figure}[h!]
\centering
\includegraphics[width=0.8\textwidth]{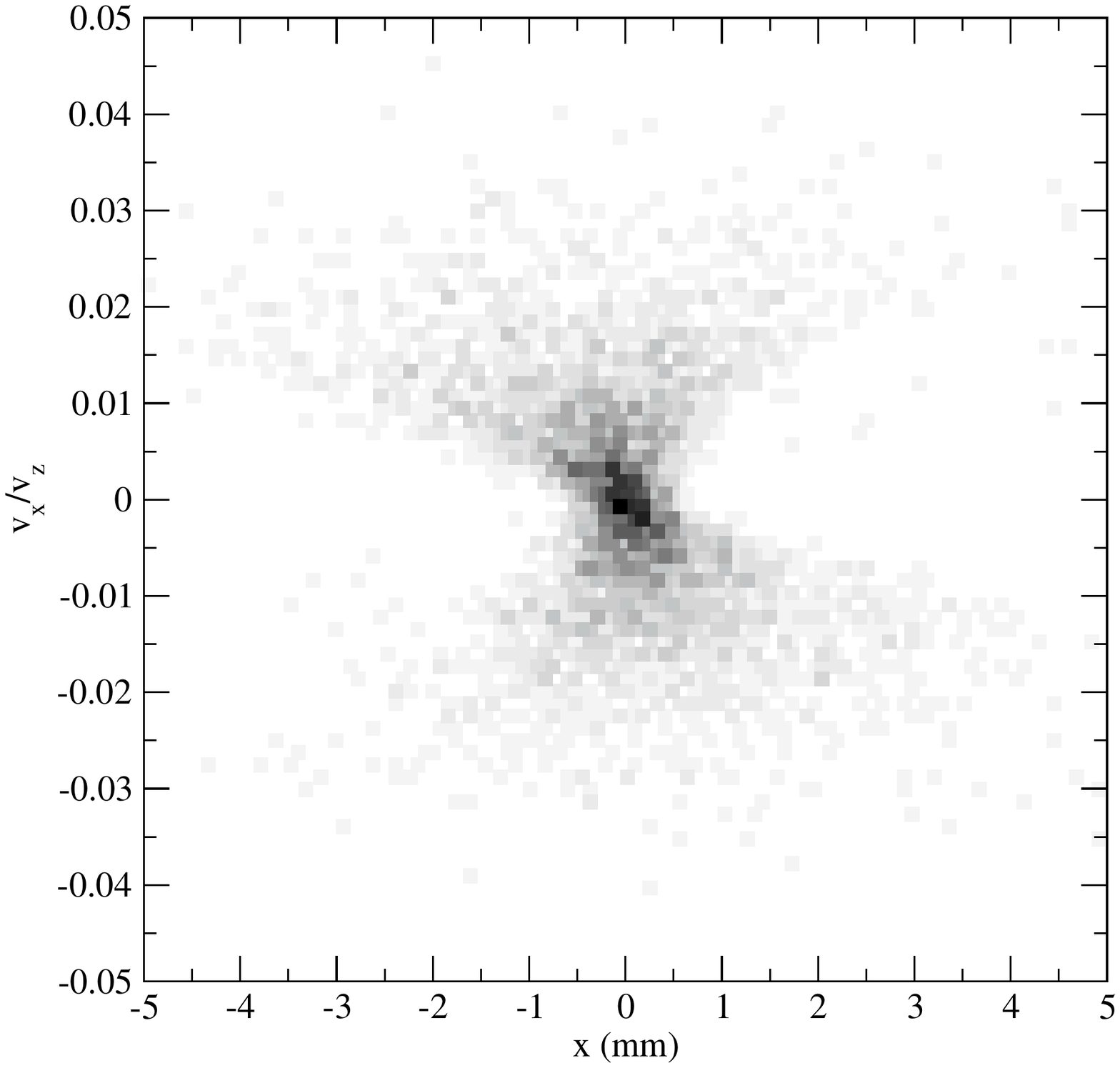}
\caption{Phase space of $\left(x, \frac{v_x}{v_z}\right)$ at simulation endpoint.}
\label{fig:phase}
\end{figure}

The $\left(x, \frac{v_x}{v_z}\right)$ phase space at the simulation exit is shown in Figure \ref{fig:phase}, illustrating a beamspot with $\sigma \approx 0.5$ mm, and emittance $\epsilon_{rms,x} = 18.5$ mm mrad. This is a beamspot comparable to the input from ELENA, but with emittance increased by around a factor of 20. These beam parameters should allow efficient capture of \Hbarplus* in a centimetre-scale Paul trap \cite{GBAR2011}.\par

\begin{figure}[h!]
\centering
\includegraphics[width=0.8\textwidth]{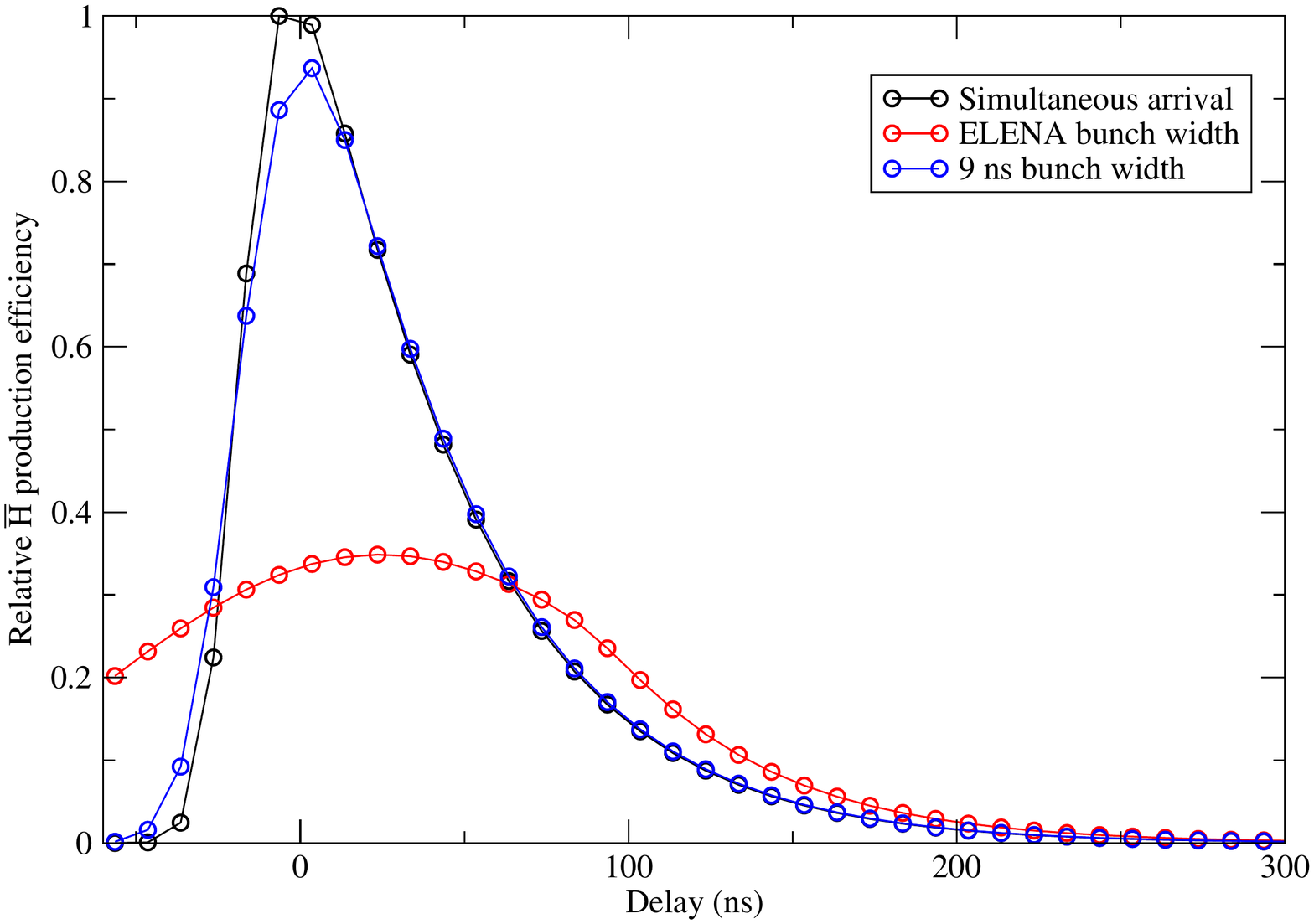}
\caption{Effect of \pbar* bunching on relative \Hbar* production efficiency, shown for the unbunched case, using the bunch width from this simulation, and for the zero bunch width case.}
\label{fig:bun}
\end{figure}

The increase in \Hbar* production with this scheme of approximately a factor of is illustrated in Figure \ref{fig:bun} (as calculated in \cite{Crivelli2016}).

\section{Conclusions}
Simulation of \pbar* time-bunching followed by energy bunching has been performed in order to simultaneously maximize at-target density while minimizing subsequent energy spread which is very important in order to efficiently capture those ions in a Paul trap as planned in the context of the GBAR experiment. This was achieved without significantly reducing the transmitted fraction of antiprotons through the target compared to the unbunched case. An increase in \Hbarplus* formation by approximately an order of magnitude could be expected by the implementation of the proposed scheme. Further improvements might be possible by using an antiproton trap which would allow for a shorter bunch owing to the smaller initial emittance assuming that the same can be achieved with the positron bunch. 

\section{Acknowledgements}
This work is supported by the SNFS under the grant number 200021\_166286.

\section{References}
\bibliography{bibliography}

\end{document}